\input{aipcheck}

\documentclass[
    ,draft            
  ]
  {aipproc}

\layoutstyle{8x11single}

\begin{document}

\title{Strangeness S=-3 and -4 baryon-baryon interactions in chiral EFT}


\author{Johann Haidenbauer}{
  address={Institute for Advanced Simulation, Institut f\"ur Kernphysik, 
and J\"ulich Center for Hadron
Physics, Forschungszentrum J\"ulich, D-52425 J\"ulich, Germany}
}

\begin{abstract}
I report on recent progress in the description of
baryon-baryon systems within chiral effective field theory.
In particular, I discuss results for the strangeness
$S=-3$ to $-4$ baryon-baryon systems, obtained to leading order. 
\end{abstract}

\maketitle


\section{Introduction}
\label{intro}

Chiral effective field theory (EFT) as proposed in the pioneering works of 
Weinberg \cite{Wei90} is a powerful tool for the derivation of nuclear forces.
In this scheme there is an underlying power counting which allows to improve calculations 
systematically by going to higher orders in a perturbative expansion. 
In addition, it is possible to derive two- and corresponding three-nucleon forces as well 
as external current operators in a consistent way. For reviews we refer the reader 
to Refs.~\cite{Bed02,Epelbaum:2005pn,Epelbaum:2008ga}. 

Over the last decade or so it has been demonstrated that 
the nucleon-nucleon ($NN$) interaction can be described to a high precision 
within the chiral EFT approach \cite{Entem:2003ft,Epe05}. Following the original
suggestion of Steven Weinberg, in these works the power counting is applied to the $NN$ 
potential rather than to the reaction amplitude. The latter is then obtained from 
solving a regularized Lippmann-Schwinger equation for the derived interaction potential. 
The $NN$ potential contains pion-exchanges and a series of contact interactions 
with an increasing number of derivatives to parameterize the shorter ranged part 
of the $NN$ force.
 
Recently, also hadronic systems involving the strange baryons $\Lambda$ and $\Sigma$, 
and the $S=-2$ baryon $\Xi$ were investigated within EFT by the group in J\"ulich
\cite{Polinder:2006zh,Polinder:2007mp,Hai07,Hai09,Hai10}. Specifically, the 
interactions in the $\Lambda N$ and $\Sigma N$ channels \cite{Polinder:2006zh} 
as well as those in the $S=-2$ sector 
($\Lambda\Lambda$, $\Sigma\Sigma$, $\Lambda\Sigma$, $\Xi N$) 
\cite{Polinder:2007mp} were considered. 
In these works the same scheme as applied in Ref.~\cite{Epe05} to the $NN$ 
interaction is adopted. 
In the present contribution I focus on a recent extension of that study to 
systems with $S=-3$ and $-4$ \cite{Hai09}.
 

\section{Formalism}
\label{sec:2}
To leading order (LO) in the power counting, as considered in the 
aforementioned investigations \cite{Polinder:2006zh,Polinder:2007mp,Hai09}, 
the baryon-baryon 
potentials involving strange baryons consist of four-baryon contact terms without 
derivatives and of one-pseudoscalar-meson exchanges, analogous to the $NN$ 
potential of \cite{Epe05}. 
The potentials are derived using constraints from ${\rm SU(3)}$ flavor symmetry. 
Details on the derivation of the chiral potentials for the $S=-1$ to $S=-4$ sectors at 
LO using the Weinberg power counting can be found in Ref. \cite{Polinder:2006zh}.
The contributions of one-pseudoscalar-meson exchanges are identical to those already 
discussed extensively in the literature, see, e.g., \cite{Polinder:2006zh}.
The LO ${\rm SU(3)}_{\rm f}$ invariant contact terms for the octet baryon-baryon interactions 
that are Hermitian and invariant under Lorentz transformations follow from the Lagrangians
\begin{eqnarray}
{\mathcal L}^1 &=& C^1_i \left<\bar{B}_a\bar{B}_b\left(\Gamma_i B\right)_b\left(\Gamma_i B\right)_a\right>\ , \quad
{\mathcal L}^2 = C^2_i \left<\bar{B}_a\left(\Gamma_i B\right)_a\bar{B}_b\left(\Gamma_i B\right)_b\right>\ , \nonumber \\
{\mathcal L}^3 &=& C^3_i \left<\bar{B}_a\left(\Gamma_i B\right)_a\right>\left<\bar{B}_b\left(\Gamma_i B\right)_b\right>\  .
\label{eq:2.1}
\end{eqnarray}
As described in \cite{Polinder:2006zh}, in LO the Lagrangians give rise to six independent
low-energy coefficients (LECs) -- $C^1_S$, $C^1_T$, $C^2_S$, $C^2_T$, $C^3_S$ and $C^3_T$ --
that need to be determined by a fit to experimental data.
Here $S$ and $T$ refer to the central and spin-spin parts of the potential, respectively.
The scattering amplitude is obtained by solving a Lippmann-Schwinger equation for the 
LO potential. Thereby, the possible coupling between different baryon-baryon channels, 
$\Lambda N-\Sigma N$ or $\Xi\Lambda-\Xi\Sigma$, say, is taken into account. 
The potentials in the LS
equation are cut off with a regulator function, $\exp\left[-\left(p'^4+p^4\right)/\Lambda^4\right]$,
in order to remove high-energy components of the baryon and pseudoscalar meson fields \cite{Epe05}.

\section{Results and discussion}
\label{sec:3}

The LO chiral EFT interaction for the $S=-3$ and $-4$ baryon-baryon sector depends only 
on those five contact terms that enter also in the $YN$ interaction, cf. Table 1
in \cite{Hai09}. 
Thus, based on the values which were fixed in our study of the $YN$ sector 
\cite{Polinder:2006zh} 
we can make genuine predictions for the interaction in the $S=-3$ and $-4$ channels 
that follow from the imposed ${\rm SU(3)_f}$ symmetry.

\begin{figure}
 \includegraphics[height=.3\textheight]{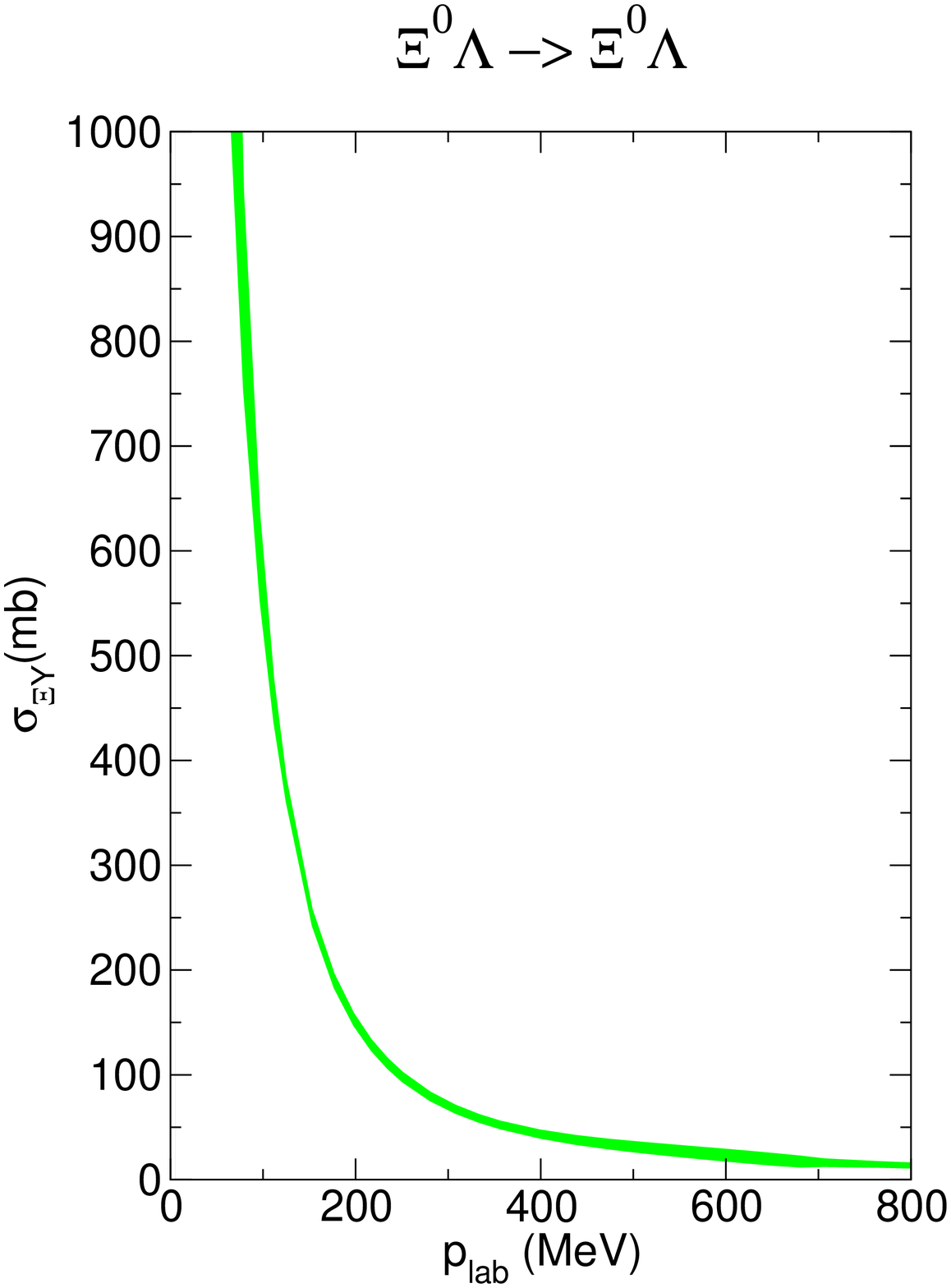}
 \includegraphics[height=.3\textheight]{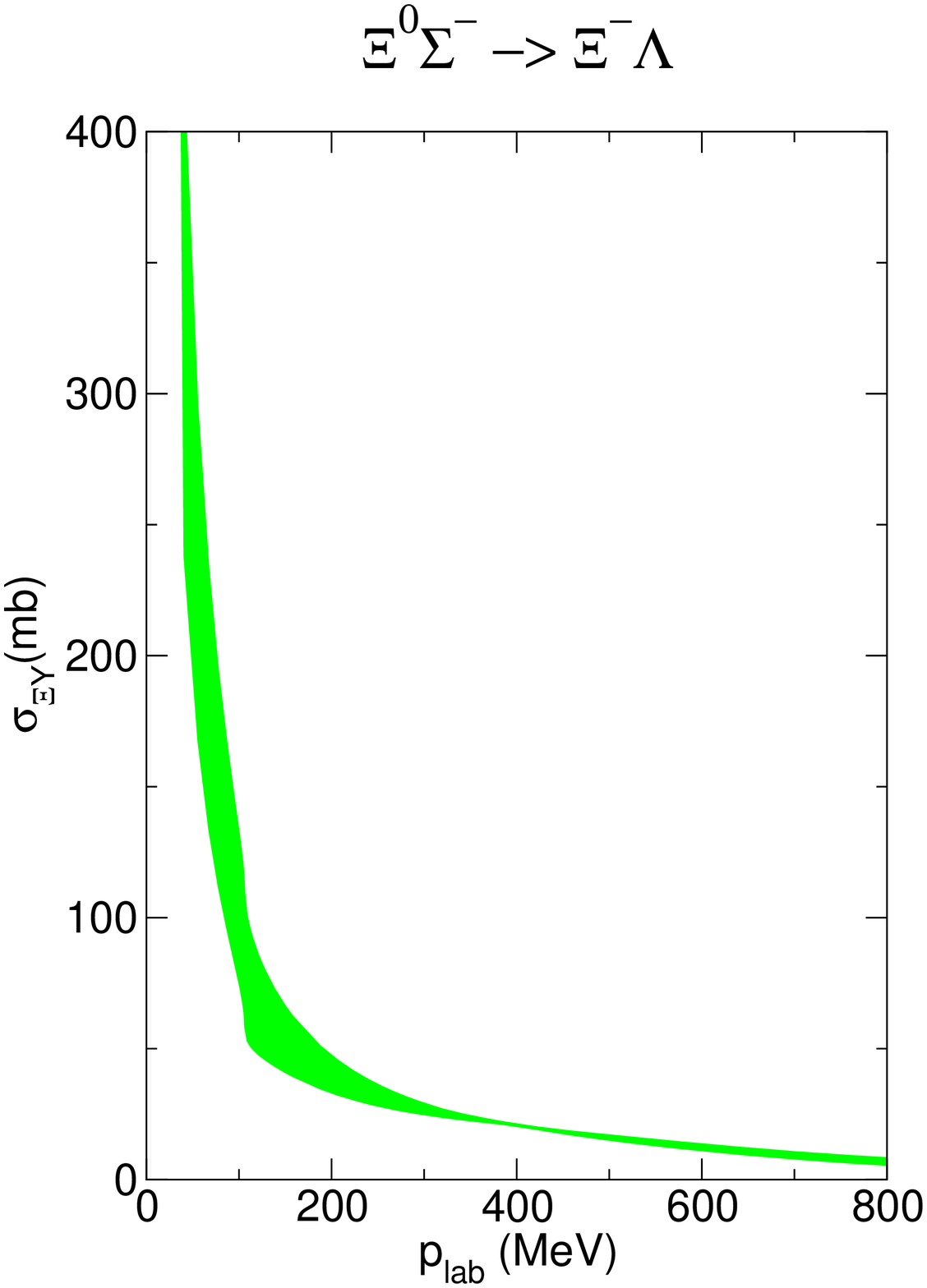}
 \includegraphics[height=.3\textheight]{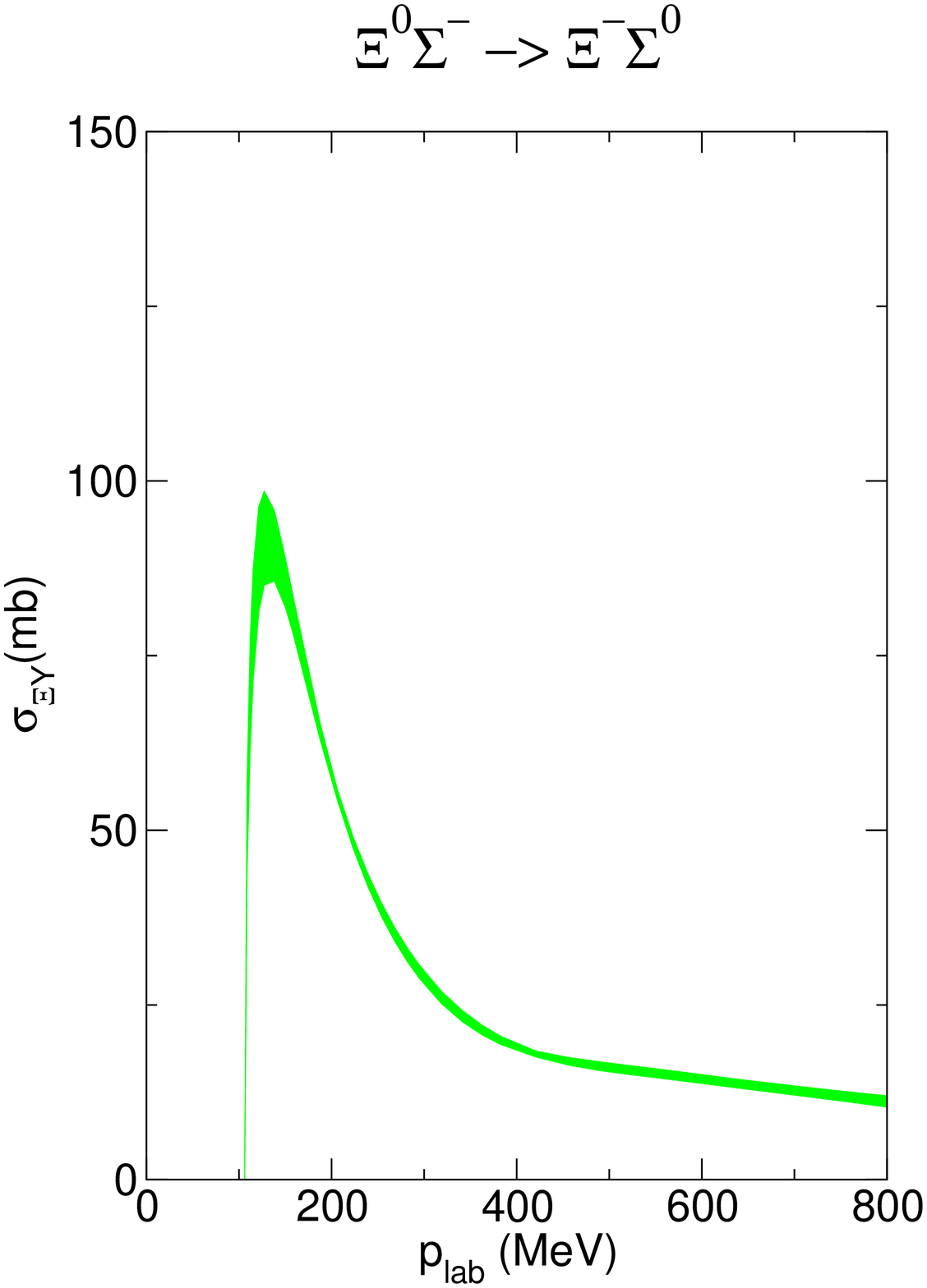} 
\end{figure}
\begin{figure}
 \includegraphics[height=.3\textheight]{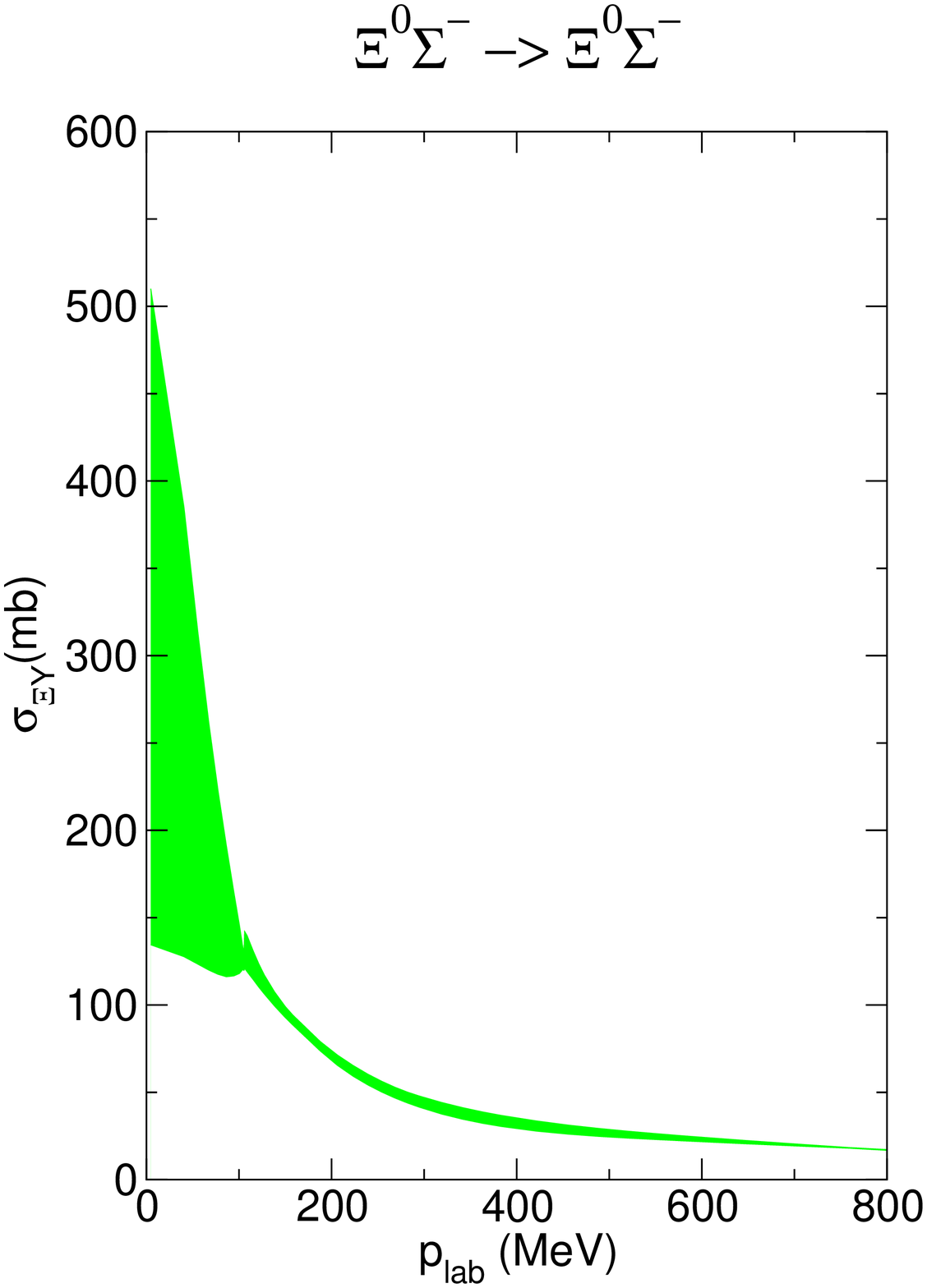}
 \includegraphics[height=.3\textheight]{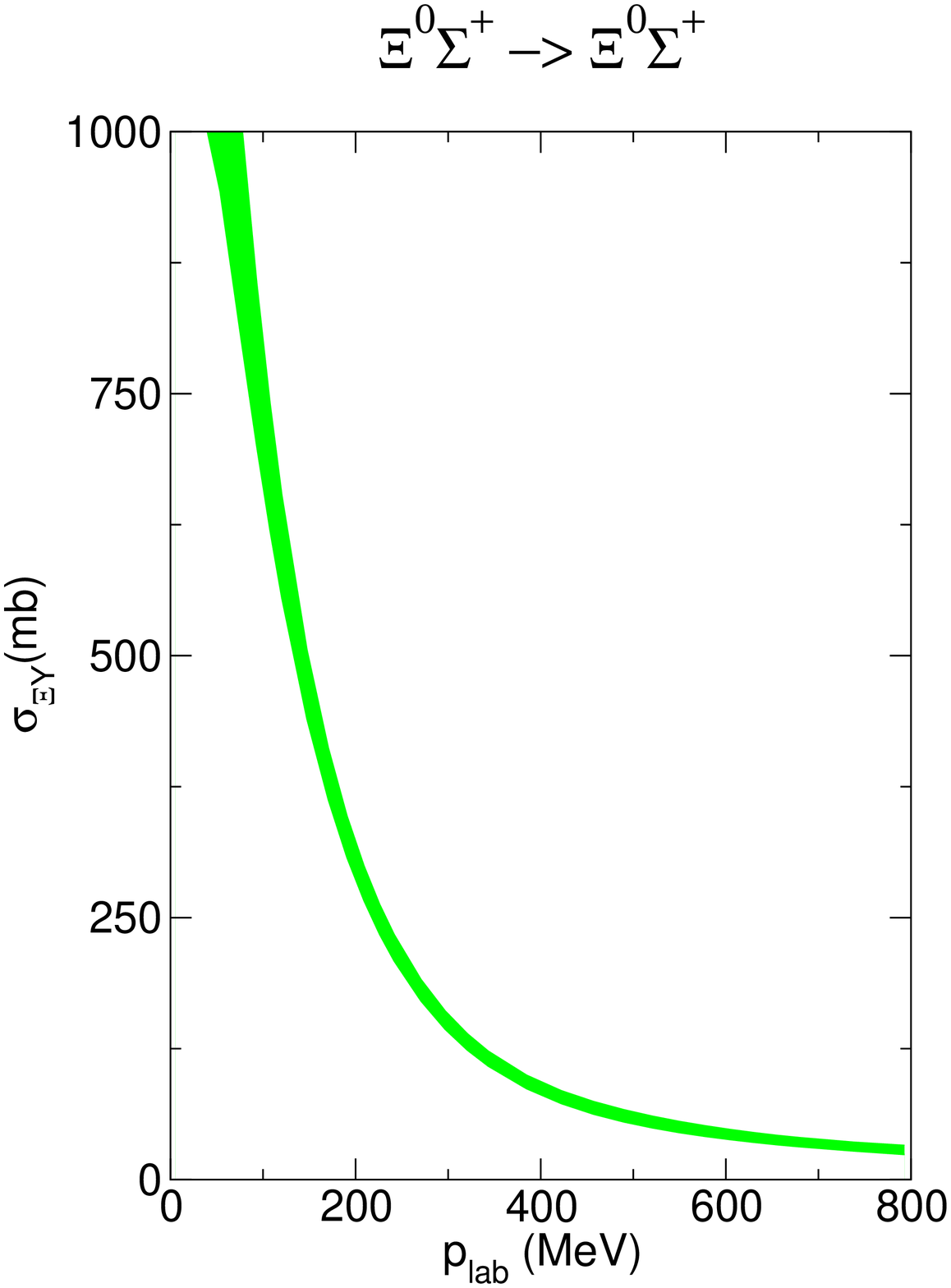}
 \caption{Total cross sections for various reactions in the strangeness $S=-3$ sector 
as a function of $p_{lab}$.
The shaded band shows the chiral EFT results for variations of the cut-off
in the range $\Lambda =$ 550$\ldots$700~MeV.
}
\label{fig:4.2}
\end{figure}

Corresponding results for the $\Xi^0\Lambda \to \Xi^0\Lambda$, $\Xi^0\Sigma^-\to \Xi^-\Lambda$,
$\Xi^0\Sigma^-\to \Xi^-\Sigma^0$, $\Xi^0\Sigma^-\to \Xi^0\Sigma^-$, and 
$\Xi^0\Sigma^+\to \Xi^0\Sigma^+$ scattering cross sections are presented 
in Fig.~\ref{fig:4.2}. Partial waves with total angular momentum 
up-to-and-including $J = 2$ are taken into account. 
The shaded bands show the cut-off dependence. 
{}From that figure one observes that the $\Xi^0\Lambda \to \Xi^0\Lambda$
and $\Xi^0\Sigma^+\to \Xi^0\Sigma^+$ cross sections are rather large near threshold. 
Though the cross section for $\Xi^0\Sigma^-\to \Xi^-\Lambda$ rises too, 
in this case it is only due to the phase space factor 
$p_{\Xi^-\Lambda}/p_{\Xi^0\Sigma^-}$. 
There is a clear cusp effect visible in 
the $\Xi^0\Sigma^-$ cross section at $p_{lab} \approx 106$ MeV/c, i.e. at
the opening of the $\Xi^-\Sigma^0$ channel. On the other hand, we do not observe
any sizeable cusp effects in the $\Xi^0\Lambda$ cross section around $p_{lab} = 690$ 
MeV/c, i.e. at the opening of the $\Xi\Sigma$ channels. The latter is in line with 
the results reported by the Nijmegen group for their interactions \cite{Stoks:1999bz},
where a cusp effect in that channel is absent too. 
In this context I would like to remind the reader that the cusp seen in the 
corresponding strangeness $S=-1$ case, namely in the 
$\Lambda N$ cross section at the $\Sigma N$ threshold,  
is rather pronounced in our chiral EFT interaction \cite{Polinder:2006zh} 
but also in conventional meson-exchange potential models \cite{Hai05,Hai07a,Rij99}.

\begin{figure}
 \includegraphics[height=.3\textheight]{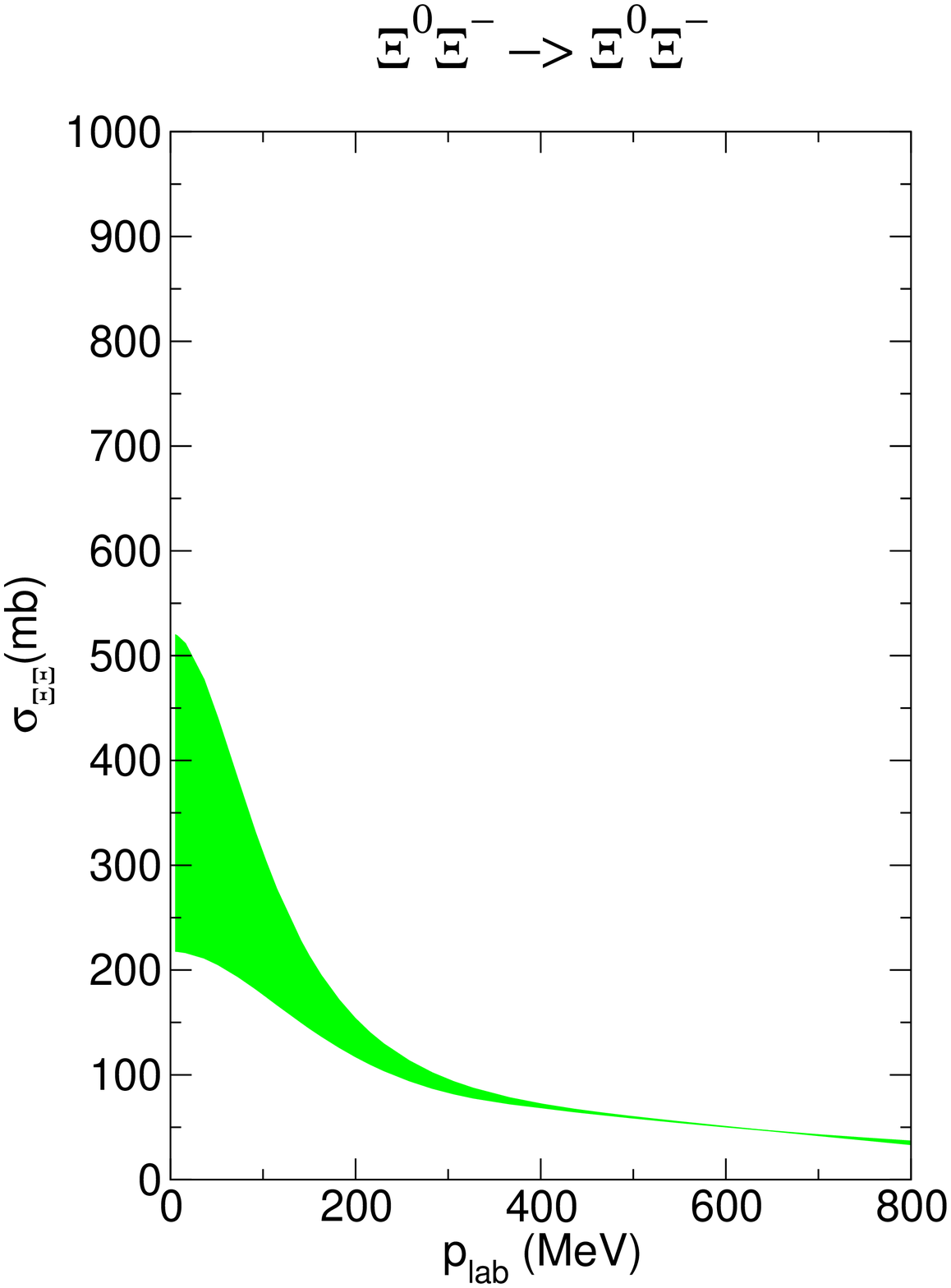}
 \includegraphics[height=.3\textheight]{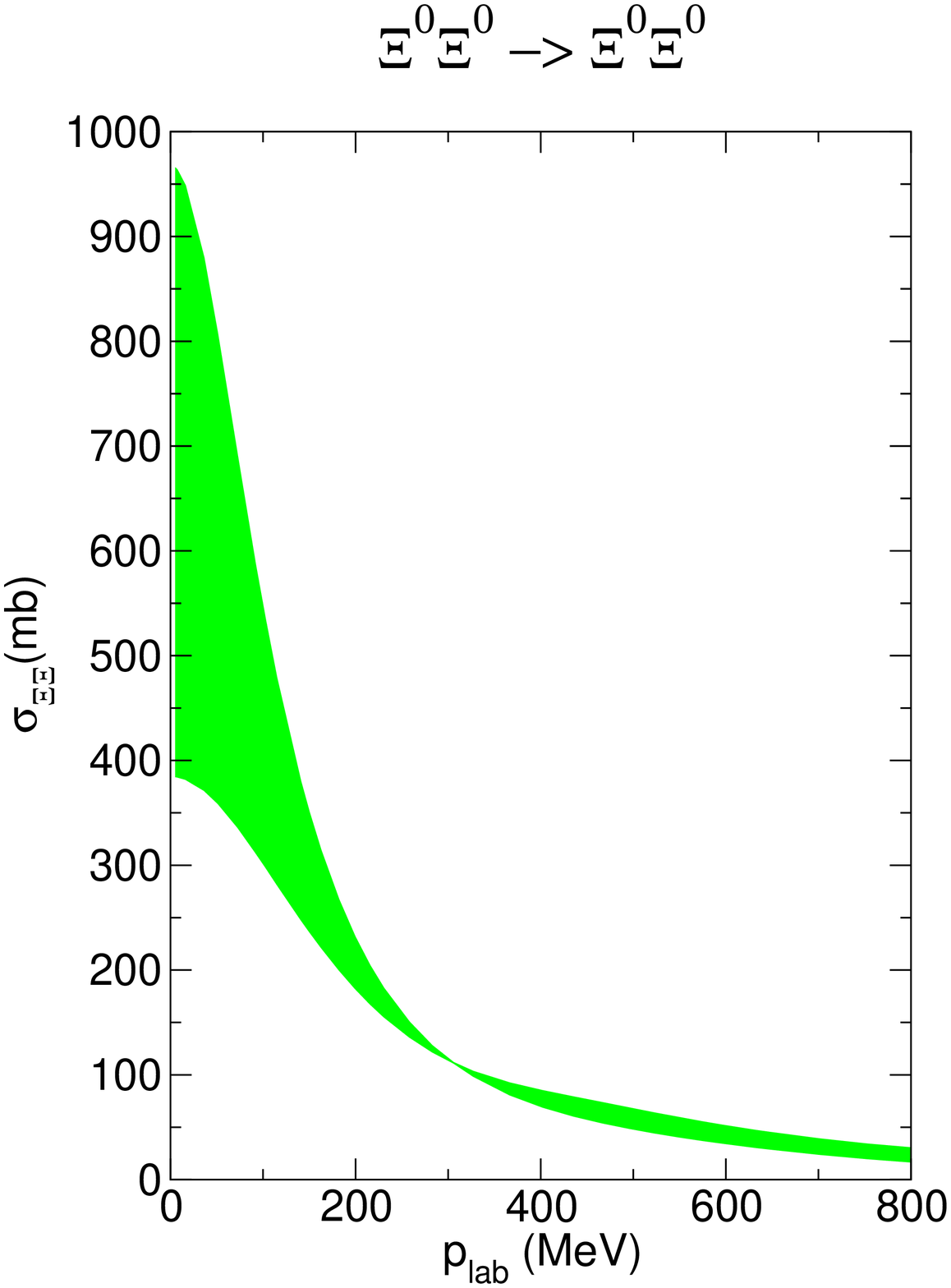}
\caption{Total cross sections for the reactions 
$\Xi^0\Xi^-\to \Xi^0\Xi^-$ and $\Xi^0\Xi^0\to \Xi^0\Xi^0$
as a function of $p_{lab}$.
The shaded band shows the chiral EFT results for variations of the cut-off
in the range $\Lambda =$ 550$\ldots$700 MeV.
}
\label{fig:4.3}
\end{figure}

Predicted cross sections for the $\Xi^0\Xi^0$ and $\Xi^0\Xi^-$ channels are shown in 
Fig.~\ref{fig:4.3}, again as a function of $p_{lab}$ and with 
shaded bands that indicate the cut-off dependence. 

\begin{table}[t]
\renewcommand{\arraystretch}{1.1}
\caption{Selected $\Xi Y$ and $\Xi\Xi$ 
singlet and triplet scattering lengths $a$ and effective ranges $r$ 
(in fm) for various cut-off values $\Lambda$. 
The last columns show results for the Nijmegen potential (NSC97a, NSC97f) \cite{Stoks:1999bz} 
and the model by Fujiwara et al. (fss2) \cite{Fujiwara:2006yh}. 
}
\label{tab:2}
\vspace{0.2cm}
\centering
\begin{tabular}{|c|cccc|cc|c|}
\hline
& \multicolumn{4}{|c|}{EFT} & \ NSC97a \ & \ NSC97f \ & \ fss2 \ \\
\hline
${\Lambda}$ (MeV)  & \ 550 \ & \ 600 \ & \ 650 \ & \ 700 \ & & & \\
\hline
$a^{\Xi\Lambda}_s$  &$-33.5$ &$35.4$ &$12.7$ &$9.07$ & $-0.80$ & $-2.11$ & $-1.08$\\
$r^{\Xi\Lambda}_s$  &$1.00$  &$0.93$  &$0.87$  &$0.84$ & $4.71$ & $3.21$ & $3.55$\\
$a^{\Xi\Lambda}_t$  &$0.33$  &$0.33$ &$0.32$ &$0.31$ & $0.54$ & $0.33$ & $0.26$\\
$r^{\Xi\Lambda}_t$  &$-0.36$  &$-0.30$  &$-0.29$  &$-0.27$ & $-0.47$ & $2.79$ &$2.15$\\
\hline
$a^{\Xi^0\Sigma^+}_s$  &$4.28$ &$3.45$  &$2.97$  &$2.74$ & $4.13$ & $2.32$ & $-4.63$\\
$r^{\Xi^0\Sigma^+}_s$  &$0.96$  &$0.90$  &$0.84$   &$0.81$ & $1.46$ & $1.17$ & $2.39$\\
$a^{\Xi^0\Sigma^+}_t$  &$-2.45$  &$-3.11$ &$-3.57$   &$-3.89$ & $3.21$ & $1.71$ & $-3.48$\\
$r^{\Xi^0\Sigma^+}_t$  &$1.84$ &$1.72$ & $1.70$ & $1.70$ & $1.28$ & $0.96$ & $2.52$\\
\hline
\hline  
$a^{\Xi\Xi}_s$  &$3.92$ &$3.16$ &$2.71$ &$2.47$ & $17.28$ & $2.38$ & $-1.43$\\
$r^{\Xi\Xi}_s$  &$0.92$  &$0.85$  &$0.79$  &$0.75$ & $1.85$ & $1.29$ & $3.20$\\
$a^{\Xi\Xi}_t$  &$0.63$ &$0.59$   &$0.55$ &$0.52$ &  $0.40$ & $0.48$ & $3.20$\\
$r^{\Xi\Xi}_t$  &$1.04$  &$1.05$  &$1.08$ &$1.11$ & $3.45$ & $2.80$ &$0.22$\\
\hline
\end{tabular}
\renewcommand{\arraystretch}{1.0}
\end{table}

Results for the $\Xi^0\Lambda$, $\Xi^0\Sigma^+$, and $\Xi\Xi$ 
scattering lengths and effective ranges are listed in Table~\ref{tab:2}. 
Here we also include predictions by other models \cite{Stoks:1999bz,Fujiwara:2006yh}
for channels where pertinent results are available in the literature.
This Table reveals the reason for the sizeable $\Xi^0\Lambda$ cross
section predicted by the chiral EFT interactions, 
namely a rather large scattering length in the corresponding 
${}^1S_0$ partial wave. It is obvious that its value is 
strongly sensitive to cut-off variations. It even changes sign (in other
words, it becomes infinite) within the considered cut-off range. This
means that a virtual bound state transforms into a real bound state,
where the strongest binding occurs for the cut-off 
$\Lambda = 700$ MeV and leads to a binding energy of $-0.43\,$MeV.
While this behaviour is interesting per se, one certainly has to stress
that in such a case the predictive power of our LO calculation is 
rather limited. One has to wait for at least an NLO calculation, where we
expect that the cut-off dependence will become much weaker so that more 
reliable conclusions on the possible existence of a virtual or a real 
bound state should be possible. 
The ${}^1S_0$ scattering lengths of the other potentials suggest also an
overall attractive interaction in this partial wave though only a very
moderate one. 

The results for the ${}^3S_1$ state of the $\Xi^0\Lambda$ channel are fairly
similar for all considered interactions. Moreover, with regard to the chiral EFT 
interaction there is very little cut-off dependence. 
The $S$-waves in the $\Xi\Sigma$ $I=3/2$ channel belong to the same 
($10^*$ and $27$, respectively) irreducible representations where in the 
$NN$ case real (${}^3S_1-{}^3D_1$) or virtual (${}^1S_0$) bound states exist,
cf. Table~1 in Ref.~\cite{Hai09}. Therefore, one expects that such states can also
occur for $\Xi\Sigma$. 
Indeed, bound states are present for both partial waves in
the Nijmegen model, cf. the discussion in Sect.~III.B in 
Ref.~\cite{Stoks:1999bz}. Their presence is reflected in the positive and
fairly large singlet and triplet scattering lengths for $\Xi^0\Sigma^+$, cf.
Table~\ref{tab:2}. The chiral EFT interaction has positive scattering 
lengths of comparable magnitude for ${}^1S_0$, for all cut-off values, and
therefore bound states, too.  These binding energies lie in the range of 
$-2.23\,$MeV ($\Lambda = 550\,$MeV) to $-6.15\,$MeV (700 MeV). 
In the ${}^3S_1-{}^3D_1$ partial wave the
attraction is obviously not strong enough to form a bound state. The same
is the case (but for both $S$ waves) for the quark model fss2 of Fujiwara et al. 
\cite{Fujiwara:2006yh}.

The ${}^1S_0$ state of the $\Xi\Xi$ channel belongs also to the $27$plet
irreducible representation and also here the Nijmegen as well as the 
chiral EFT interactions produce bound states. In our case the binding energies
lie in the range of $-2.56\,$MeV ($\Lambda = 550\,$MeV) to $-7.28\,$MeV 
(700 MeV). The predictions of both approaches for the ${}^3S_1$ scattering 
length are comparable. 
The quark model of Fujiwara et al. exhibits a different 
behavior for the $\Xi\Xi$ channel, see the last column in 
Table~\ref{tab:2}. The small and negative ${}^1S_0$ 
scattering length signals an interaction that is only moderately attractive.
The large and positive scattering length in the ${}^3S_1-{}^3D_1$ 
partial wave, produced by that potential model, is usually a sign 
for the presence of a bound state, though according to the authors this 
is not the case for this specific interaction. 
Further results, and specifically $\Xi\Lambda$, $\Xi\Sigma$, and $\Xi\Xi$
phase shifts, can be found in \cite{Hai10a}.

\section{Summary}
\label{chap:5}

Our investigations show that the chiral EFT scheme, successfully applied in Ref.~\cite{Epe05} 
to the $NN$ interaction, also works well for the $\Lambda N$, $\Sigma N$ 
\cite{Polinder:2006zh,Hai10} and $\Lambda \Lambda$ \cite{Polinder:2007mp}
interactions. Moreover, as reported here, it can be used o make predictions for
the $S=-3$ and $-4$ baryon-baryon interactions invoking constraints from
${\rm SU(3)}$ flavor symmetry. 
It will be interesting to see whether the new facilities J-PARC (Tokai, Japan) and
FAIR (Darmstadt, Germany) allow access to empirical information about the
interaction in the $S=-3$ and $-4$ sectors. Such information could come from
formation experiments of corresponding hypernuclei or from proton-proton and
antiproton-proton collisions at such high energies that pairs of baryons with
strangeness $S=-3$ or $S=-4$ can be produced.


\end{document}